\documentclass[%
 reprint,
superscriptaddress,
 amsmath,amssymb,
 aps,
]{revtex4-2}

\usepackage{graphicx}
\usepackage{dcolumn}
\usepackage{bm}
\usepackage{color} 
\usepackage{here}
\usepackage{comment}
\usepackage{ulem}

\begin{document}

\preprint{APS/123-QED}

\title{On the Novel Superfluidity in the Second Layer of $^4$He on Graphite}

\author{Jun Usami}
 \email{j-usami@aist.go.jp}
\affiliation{Cryogenic Research Center, The University of Tokyo, 2-11-16 Yayoi, Bunkyo-ku, Tokyo 113-0032, Japan\looseness=-1}
\affiliation{Department of Physics, The University of Tokyo, 7-3-1 Hongo, Bunkyo-ku, Tokyo 113-0033, Japan\looseness=-1}
\affiliation{National Institute of Advanced Industrial Science and Technology (AIST), 1-1-1 Higashi, Tsukuba, Ibaraki 305-8565, Japan\looseness=-1}
\author{Hiroshi Fukuyama}%
 \email{hiroshi@kelvin.phys.s.u-tokyo.ac.jp}
\affiliation{Cryogenic Research Center, The University of Tokyo, 2-11-16 Yayoi, Bunkyo-ku, Tokyo 113-0032, Japan\looseness=-1}
\affiliation{Department of Physics, The University of Tokyo, 7-3-1 Hongo, Bunkyo-ku, Tokyo 113-0033, Japan\looseness=-1}

\date{\today}

\begin{abstract}

Evidence for a new type of superfluid phase in second-layer $^4$He on graphite has been obtained from simultaneous measurements of torsional-oscillator response and heat-capacity on exactly the same sample down to 30 mK, which resolve substrate-related uncertainties in the previous studies.
The new phase hosts both superfluidity and enhanced viscoelasticity, and is stable over a finite density interval, strongly supporting the proposed {\it superfluid liquid-crystal} hypothesis.
A random-Josephson-network analysis shows that the widely reported log-$T$ dependence of the superfluid density is likely due to substrate imperfections.

\end{abstract}

\maketitle



%
\begin{figure}[b]
   \centering
    \includegraphics[width=1.00\linewidth]{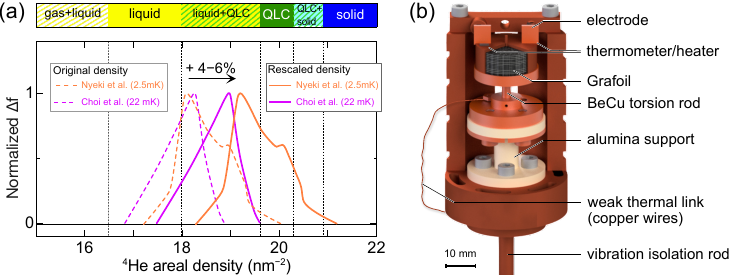} 
      \caption{(a) Normalized TO superfluid responses in the second layer of $^4$He on graphite (Grafoil) (obtained from Refs.~\cite{Nyeki2017a,Choi2021}) plotted using each group's original density axis (dashed curves): the density for each group (solid curves) was rescaled by the respective authors themselves to conform to the thermodynamic phase diagram determined by the HC measurement~\cite{Nakamura2016} (shown at the top). However, no established protocol exists to perform the rescaling. (b) Experimental apparatus specifically designed to measure TO signals and HCs simultaneously on the same He sample~\cite{Usami2022}.}
   \label{fig-setup}
\end{figure}

Superfluidity and superconductivity are characterized by phase coherence over macroscopic length scales, usually resulting in a spatially uniform distribution of constituent particles.
However, in some cases, the off-diagonal long-range order (LRO) of superfluidity and superconductivity can coexist with {\it spontaneous} density modulations that are diagonal LROs.
One prominent example is the so-called supersolid, where superfluidity coexists with crystalline order.
This possibility has long been investigated theoretically~\cite{Andreev1969,Chester1970,Leggett1970} and experimentally in bulk solid $^{4}$He~\cite{Kim2004a}, but so far, its emergence has been indicated only in the presence of an extrinsic disorder~\cite{Chan2013,Beamish2020,Kuklov2025}.
More broadly, superfluid liquid-crystals with partially broken spatial symmetries~\cite{Beekman2017} are under active investigation across diverse quantum systems, including dilute cold gases~\cite{Tanzi2019,Bottcher2021,Norcia2021}, correlated electrons~\cite{Fradkin2015}, and nuclear matter~\cite{Caplan2017}.

There are promising arguments about the possible emergence of superfluid liquid-crystal or even supersolid phases for the second monolayer of bilayer $^{4}$He physisorbed on a graphite surface~\cite{Nakamura2016,Nyeki2017a,Nyeki2017,Choi2021,Knapp2025}.
This is a structurally simple and chemically pure two-dimensional (2D) bosonic system, thus an intuitive candidate for those exotic states.
Here, particle correlations are highly tunable by varying the $^{4}$He areal density ($\rho$).
Ref.~\cite{Nakamura2016} reports that, regardless of system size, a distinct intervening phase, which is absent in the three-dimensional He system, exists in between the quantum liquid and solid phases over a finite density span (0.7\,nm$^{-2}$ wide), as shown at the top of Fig.\,\ref{fig-setup}.
Such a detailed phase diagram became available through high-precision heat-capacity (HC) measurements of melting anomalies using a high-quality graphite substrate~\cite{Nakamura2016}, since HC is quite sensitive to phase transitions as functions of  temperature and density. 
A more recent experiment~\cite{Knapp2025}, where the $^{3}$He impuriton state was observed over an extended $^{4}$He density range of order of 1\,nm$^{-2}$, seems to be consistent with this phase diagram.
The intervening phase is considered to be a quantum liquid crystal (QLC), more specifically the quantum hexatic~\cite{Nakamura2016} or the similar density modulated phase~\cite{Nyeki2017a}.
At densities near the QLC phase, increases in the resonance frequency ($\Delta f$) of a torsional-oscillator (TO) are observed at temperatures lower than $200$\,mK by Crowell and Reppy~\cite{Crowell1993,Crowell1996}, and later, by other groups~\cite{Shibayama2009,Nyeki2017a,Choi2021}. 
$\Delta f$ is suggestive of nonclassical rotational inertia (NCRI), a fingerprint of superfluidity~\cite{Leggett1970}.
Superfluidity was further supported by observations of the frequency-, velocity-, and $^{3}$He impurity-independence of $\Delta f$~\cite{Choi2021} and by its sizable magnitude (0.8$\rho$), measured down to $2.5$\,mK~\cite{Nyeki2017a}.
A quantum Monte Carlo calculation~\cite{Gordillo2020} predicts a supersolid phase with a reduced superfluid density ($\approx0.3\rho_{\mathrm{2nd}}$) at a density near the QLC phase.
Here, $\rho_{\mathrm{2nd}}$ is the $^4$He areal density in the second layer.

However, the situation is still far from a complete understanding mainly because of the following three issues.
(i) One is the lack of direct experimental information on the structure of the QLC phase from scattering experiments because of the extreme temperature environment~\cite{Carneiro1981,Lauter1990,Lauter1992}.
(ii) The second is that exfoliated graphite substrates (typically Grafoil), which are commonly used in TO measurements, have certain amounts (10--15\%) of surface heterogeneities~\cite{Godfrin1995,Sato2012,Saunders2020} because of their platelet (microcrystallite) structure~\cite{Birgeneau1982,Niimi2006}.
This results in non-negligible uncertainties in the density scales of different groups as shown in Fig.\,\ref{fig-setup}(a), when identifying which phase has what kind of superfluidity by comparing with the calorimetric (thermodynamic) phase diagram~\cite{Nakamura2016}.
Each group has calibrated their density scale at one or two characteristic densities, e.g., the submonolayer $\sqrt{3}\times\sqrt{3}$ commensurate phase formation~\cite{Shibayama2009,Choi2021}, the second-layer promotion~\cite{Choi2021}, the third-layer promotion~\cite{Nyeki2017a,Choi2021}, or the fourth-layer completion~\cite{Crowell1996}, depending on the preferred technique.
Although these attempts have been partially successful~\cite{Choi2021,Knapp2025}, the inconsistency or ambiguity in the density scale is still too large (4--6\%; see Fig.\,\ref{fig-setup}(a)), making it difficult to determine whether the QLC phase {\it itself} exhibits superfluidity.
Note that the phase diagram is quite complicated in the relevant region as seen in Fig.\,\ref{fig-setup}(a), where five different phases appear within a narrow density window and the QLC phase spans only 3.6\% in density.
A related controversial question is whether even the pure solid phase in the second layer exhibits superfluidity, i.e., the so-called supersolidity.
One of the previous TO experiments~\cite{Nyeki2017} suggested yes, while the other~\cite{Crowell1996} and the theories~\cite{Corboz2008,Gordillo2020} did not.
(iii) Lastly, all previous TO experiments reported no clear transition temperature, and mostly a curious logarithmic $T$-dependence of $\Delta f$ instead.
This is in marked contrast to a standard 2D superfluid transition with a $\Delta f$ jump at a well-defined topological phase transition temperature ($T_{\rm{BKT}}$), the Berezinskii-Kosterlitz-Thouless (BKT) transition~\cite{Berezinskii1971,Kosterlitz1973}, as seen in $^{4}$He films on Mylar~\cite{Bishop1980}.
It is still an open question whether this is intrinsic~\cite{Nyeki2017a} or extrinsic~\cite{Crowell1996}.

In this work, we have overcome the problem of density-scale ambiguity, issue (ii), by performing simultaneous measurements of TO and HC on \textit{exactly the same} 2D $^4$He (boson) and $^3$He (fermion) samples, down to 30\,mK.
The simultaneous measurements have distinct advantages as follows.
First, the most reliable and detailed phase diagram currently available has been established by HC measurements~\cite{Nakamura2016}. 
Second, HC data from different groups can be compared with exceptionally high accuracy ($\pm$0.5\% or $\pm$0.1\,nm$^{-2}$~\cite{sm}). 
These features allow us to determine the density and therefore the phase of each TO sample precisely, making it possible to identify which phases exhibit which type of superfluidity.
As a result, we detected a finite superfluid response not only in the liquid phase but also in the adjacent QLC phase for $^4$He, whereas no such response was detected for $^3$He. 
We also observed that the TO acquires an additional ``rigidity''  once the liquid sample enters the coexistence-like region (hereafter simply referred to as the ``coexistence region'') with the QLC phase~\cite{Nakamura2016}, for both $^4$He and $^3$He.
These observations strongly support the superfluid liquid-crystal scenario~\cite{Nakamura2016}.
Regarding issue (iii), we observed the log-$T$ dependence of $\Delta f$ without a clear transition temperature in both the liquid and QLC phases, suggesting that its origin is not intrinsic to the nature of $^4$He quantum state. 
Analyses based on a simple random-Josephson-network model indicate that the anomalous behaviors can arise from percolation of small superfluid islands segmented by the microcrystalline structure of the substrate.

\begin{figure}[b]
   \centering
   \includegraphics[width=0.95\linewidth]{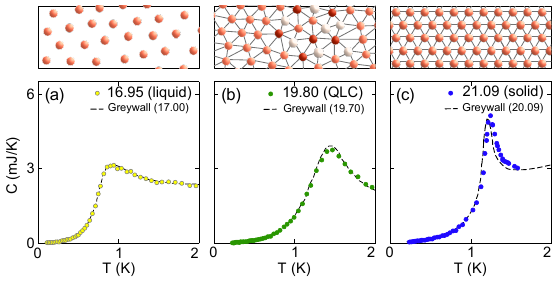} 
      \caption{HC data at densities corresponding to (a) liquid (16.95), (b) QLC (19.80), and (c) solid (21.09) phases obtained in this work (dots). They show characteristic melting anomalies with different sharpness centered at different temperatures. Close agreement is observed with the HC data obtained at nearby densities by Greywall~\cite{Greywall1993} using a Grafoil substrate (dashed curves). All numbers denote total $^4$He densities in units of nm$^{-2}$. Atomic structures of the three phases are illustrated by cartoons at the top of each panel.}
   \label{fig-HC}
\end{figure}

In the present work, we used the sample cell shown in Fig.\,\ref{fig-setup}(b), which contains a Grafoil substrate with a surface area of 44.3\,m$^{2}$~\cite{sm}.
It is mechanically supported by an alumina thermal insulation rod, and is thermally connected to the mixing chamber of a dry dilution refrigerator via a thin copper wire.
This configuration enables simultaneous TO and HC measurements, but it limits the lowest measurement temperature to 30\,mK.
The resonance frequency and the quality factor ($Q$) of this TO were $1392$\,Hz and $5\times10^{5}$, respectively, at low temperatures.
The design and performance details of this apparatus have been reported elsewhere~\cite{Usami2022}.

\begin{figure*}[t]
		\centering
			\includegraphics[width=0.85\linewidth]{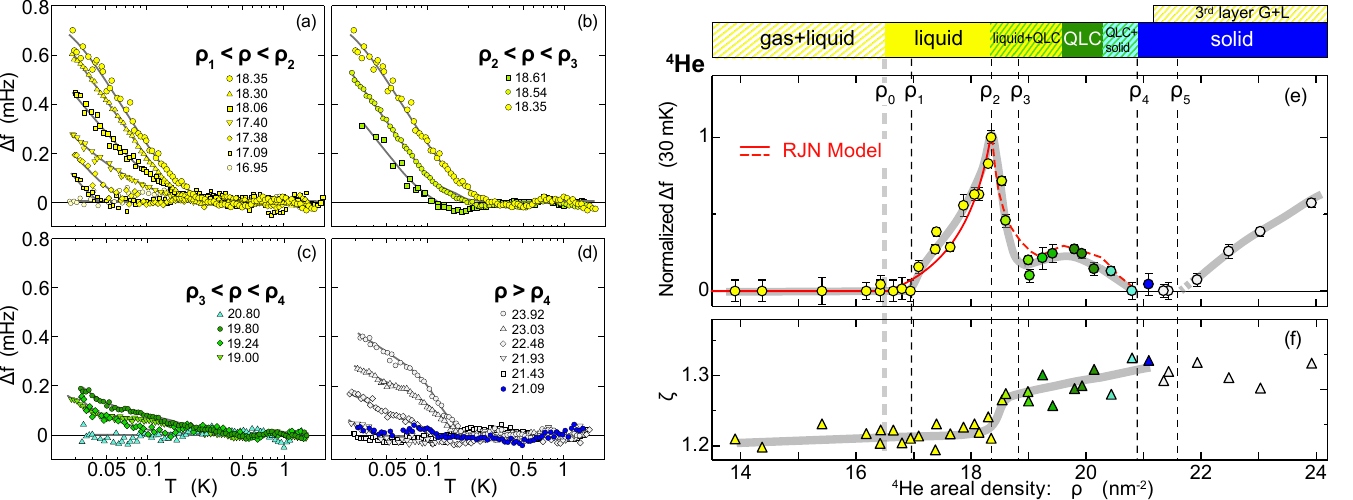}
\caption{Frequency shifts $\Delta f$ associated with superfluid NCRI for coverages between (a) 16.95--18.35, (b) 18.35--18.61, (c) 19.00--20.80, and (d) 21.09--23.92\,nm$^{-2}$.
(e) Plot of $\Delta f$ at $T = 30$\,mK vs. $^4$He density shows a systematic change at characteristic densities $\rho_1$--$\rho_5$ indicated by the vertical dashed lines, along with the thermodynamic phase diagram~\cite{Nakamura2016} shown at the top. 
Note that the low-density bound of the liquid+QLC coexistence has been set to $\rho_2$ (the same hereinafter). 
Here, $\rho_1$, $\rho_2$, $\rho_3$, $\rho_4$, and $\rho_5$ are 16.95, 18.35, 18.8, 20.9, and 21.6\,nm$^{-2}$, respectively.
$\rho_0 = 16.5$\,nm$^{-2}$ is the high-density bound of the gas+liquid coexistence region.
The thick grey curve is a guide for the eyes.
The red solid curve ($\rho \leq \rho_{2}$) and the red dashed curve ($\rho \geq \rho_{2}$) are calculated density dependences based on the RJN model (see text).
(f) Fitting parameter $\zeta$ in Eq.\,\eqref{eq-1} representing the stiffness of $^{4}$He film shows a stepwise increase at $\rho_2$ and above.}
   \label{fig-Tvsdf}
\end{figure*}

Figures~\ref{fig-HC}(a)-(c) show HC data (dots) obtained in the present work at three densities: (a) 16.95, (b) 19.80, and (c) 21.09\,nm$^{-2}$, which are representative of the (a) liquid, (b) QLC, and (c) incommensurate solid (hereafter, solid) phases, respectively. 
Corresponding adatom structures are illustrated at the top of each panel.
Each phase has its own characteristic HC peak profile, indicating the different melting mechanisms via pair unbinding of distinct topological defects: vortices (superfluid), disclinations (in the case of  hexatic QLC), and disclinations/dislocations (solid).
Our data agree remarkably well with the HC data of Greywall~\cite{Greywall1993} (the dashed curves in (a)--(c)), who also used a Grafoil substrate, and with those of Nakamura \textit{et~al}.~\cite{Nakamura2016}, who used a ZYX substrate, over the entire density range studied here, as demonstrated in Fig.\,S1~\cite{sm}.
Therefore, HC is an ideal navigator for phase assignment.

Figures~\ref{fig-Tvsdf}(a)--(d) show the temperature dependences of our $\Delta f$ data plotted as functions of $\log T$, and Fig.\,\ref{fig-Tvsdf}(e) shows the $\rho$-dependence of the $\Delta f$ at a fixed temperature ($30$\,mK).
As $\rho$ increases from 14\,nm$^{-2}$, we first detected a finite $\Delta f$ at 17.09\,nm$^{-2}$ (we define the last non-superfluid density 16.95\,nm$^{-2}$ as $\rho_1$; see (e)) below a certain onset temperature, $T_{\rm{onset}} \approx 50$\,mK.
As $\rho$ increases further, $\Delta f$\,(30\,mK) and $T_{\rm{onset}}$ continuously increase up to 18.35\,nm$^{-2}$ ($\rho_2$).
Above $\rho_2$, both $\Delta f$\,(30\,mK) and $T_{\rm{onset}}$ decrease sharply.
In this density region ($\rho_1 \leq \rho \leq \rho_2$), $\Delta f$ increases logarithmically with decreasing temperature, i.e., $\Delta f \propto a \log (T/T_{\rm{onset}})$, as seen in (a), which is qualitatively similar to those reported in the previous works~\cite{Crowell1996,Nyeki2017a,Choi2021}. 
As in the previous works, the dissipation signal $Q^{-1}$ does not show any peaks in the whole temperature range, unlike the BKT transition (not shown here).

In Ref.~\cite{Nakamura2016}, the low-density bound of the liquid+QLC coexistence region was not precisely determined, but it is certainly between 18.10 and 18.70\,nm$^{-2}$.
Therefore, we conclude that the behaviour in $\rho_1 \leq \rho \leq \rho_2$ corresponds to the pure liquid phase, and the sharp peak in $\Delta f$ at $\rho_2$ marks the onset of the liquid+QLC coexistence region, providing an invaluable density-scale calibration point (see below).
We also remark that the superfluid response does not appear just above the second-layer promotion density ($\rho_{\rm{1st}\rightarrow\rm{2nd}} = 11.8\pm0.3$\,nm$^{-2}$)~\cite{Nakamura2016}, but only after the gas+liquid coexistence ends at $\rho_0 = 16.5\pm0.6$\,nm$^{-2}$.
Here we estimated the $\rho_0$ value by adding the self-bound liquid density in the second layer ($4.71\pm0.35$)\,nm$^{-2}$~\cite{Ceperley1989,Pierce1999a,Gordillo2020} to $\rho_{\rm{1st}\rightarrow\rm{2nd}}$.

$\Delta f$\,(30\,mK) ends up its rapid decrease above $\rho_2$ at 18.8\,nm$^{-2}$ ($\rho_3$), the middle density within the liquid+QLC coexistence, and then stays nearly constant at a small but finite value (approximately one quarter of the peak value at $\rho_2$) even within the $\it{pure}$ QLC phase.
$T_{\rm{onset}}$ in the QLC phase is 0.4--0.5\,K.
This is the first direct evidence that superfluidity is an intrinsic property of the QLC phase itself.
Careful analyses of the background temperature dependence of our TO, which are crucial for this conclusion, are discussed below.

The finite $\Delta f$\,(30\,mK) gradually decreases in the QLC+solid coexistence region and disappears near the end of the coexistence at 20.9\,nm$^{-2}$ ($\rho_4$).
We did $\textit{not}$ observe a finite $\Delta f$ in the pure solid phase (21.09 and 21.43\,nm$^{-2}$) within the experimental accuracy ($\pm 0.04$\,mHz) down to our lowest temperature of $30$\,mK.
In other words, we observed no trace of supersolidity in our experiment.

Above 21.6\,nm$^{-2}$ ($\rho_5$), the system again exhibits a superfluid response, where $\Delta f$\,(30\,mK) increases roughly linearly with $\rho$ (Fig.\,\ref{fig-Tvsdf}(d)).
This response likely originates from superfluid puddles in the third layer of $^4$He, considering that the previous HC experiment~\cite{Greywall1993} and theoretical calculation~\cite{Pierce1999a} indicate that the layer promotion occurs at 21.2--21.6\,nm$^{-2}$ ($\rho_{\rm{2nd}\rightarrow\rm{3rd}}$). 
Note that the density lag between $\rho_{\rm{2nd}\rightarrow\rm{3rd}}$ and $\rho_5$ in the third layer is only 0--0.4\,nm$^{-2}$.

\begin{figure}[t]
	\centering
		\includegraphics[width=1.00\linewidth]{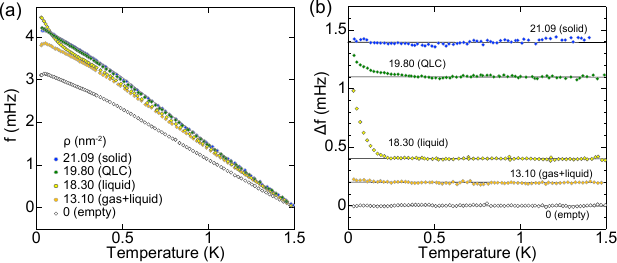} 
			\caption{(a) Temperature dependences of measured resonant-frequency raw data $f(T)$ before subtracting the substrate-He composite background $f_{\rm{cBG}} (T, \rho)$, for various He sample densities. The numbers denote total densities in atoms/nm$^2$. The data are vertically shifted to coincide at 1.5\,K. The small but rather sharp decrease in $f_{\rm{emp}}$ below 50\,mK  is characteristic of BeCu torsion rods~\cite{Agnolet1989}, and has properly been included in the $T$-dependence of $f_{\rm{emp}}$. (b) Resonance frequency shifts $\Delta f$ after subtracting $f_{\rm{cBG}} (T, \rho)$ determined by fitting the raw data shown in (a) to Eq.\,\eqref{eq-1} in the temperature range between $0.5$ and $1.5$\,K. Upturns below 0.4--0.5\,K in the liquid and QLC phases are due to superfluid mass decoupling. Each dataset is vertically shifted for clarity.}
	\label{fig-comp}
\end{figure}

We now turn to the background temperature dependence of TO, $f_{\rm{cBG}}$, which must be correctly subtracted from the raw data in order to extract $\Delta f(T)$ that reflects only superfluidity in $^{4}$He films.
Usually $f_{\rm{cBG}}$ depends not only on temperature but also on the sample density $\rho$ to a certain extent.
This is because the presence of He modifies the shear modulus of the combined system of the graphite substrate and TO, i.e., the substrate-sample composite (or viscoelastic) effect~\cite{Maris2012,Makiuchi2018}.
This effect was observed in our raw data, too, as shown in Fig.\,\ref{fig-comp}(a).  
When the composite effect is strong, a nontrivial $T$- and $\rho$-dependent background subtraction becomes necessary~\cite{Nyeki2017a}.
However, we found that, if the effect is moderate (small but finite) as a result of careful cell-designing~\cite{Usami2022}, the following simple equation can represent $f_{\rm{cBG}} (T, \rho)$ very well over a wide temperature range (0.5 $\leq T \leq 1.5$\,K) for all the sample densities investigated.

\begin{equation}
 f_{\rm{cBG}} (T, \rho) = \zeta (\rho) f_{\rm{emp}} (T) + f (T_{0}, \rho).
 \label{eq-1}
\end{equation}
Here $\zeta (\rho)$ and $f (T_{0}, \rho)$ depend only on $\rho$, and $T_{0}$ is an arbitrary reference temperature higher than $T_{\rm{onset}}$.
In our analyses, $T_{0} = 1.0$\,K where $f (T_{0}, \rho)$ is determined from the He mass loading data.
$\zeta$ represents the strength of the composite effect ($\zeta = 1$ at $\rho = 0$; empty cell), reflecting viscoelasticity (or stiffness) of He samples.

In the $\Delta f$ results shown in Fig.\,\ref{fig-Tvsdf}, $f_{\rm{cBG}}$ has already been subtracted. 
Figure\,\ref{fig-comp}(a) shows representative raw data before the background subtraction, and (b) represents corresponding results after subtraction, which are the same as those already shown in Figs.\,\ref{fig-Tvsdf}(a)--(d) but replotted as functions of linear $T$.
It reveals the excellent applicability of Eq.\,\eqref{eq-1} to our data and clear low-temperature upturns in $\Delta f$ associated with the superfluid signals in the liquid and QLC phases (Fig.\,\ref{fig-comp}(b)).
We emphasize that, unless $T_{\rm{onset}} < T_{\rm{low}} \ll T_{\rm{high}}$, it would be difficult to extract the subtle superfluid signal in the QLC phase reliably.
This condition has not necessarily been met in the previous TO experiments~\cite{Nyeki2017,Choi2021}.
Here $T_{\rm{low}}$ ($T_{\rm{high}}$) is the low (high) temperature bound of the background-data fitting to Eq.\,\eqref{eq-1}.
See Ref.~\cite{sm} for further details on the composite background.

\begin{figure}[t]
		\centering
			\includegraphics[width=0.90\linewidth]{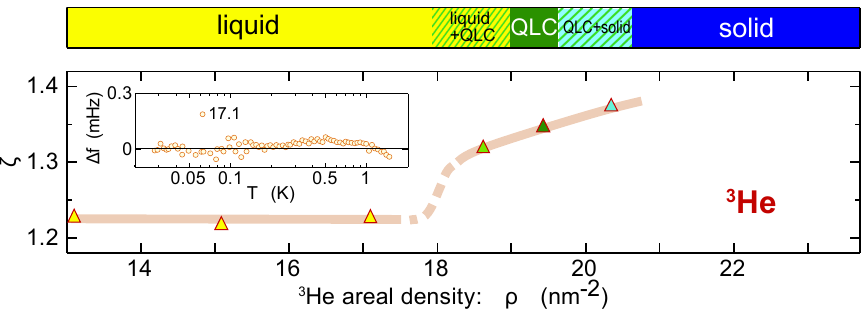}
\caption{Fitted $\zeta$ parameter values for the second layer of bilayer $^3$He on graphite. $\zeta$ shows a stepwise increase near the phase boundary ($\approx 18.0$\,nm$^{-2}$) between the liquid and the liquid+QLC coexistence region. Note that, due to different zero-point energies, the thermodynamic phase diagram of $^3$He~\cite{Nakamura2016}, shown at the top, is slightly shifted to lower densities compared with $^4$He. (Inset) Like the other five densities, no superfluid signal is observed in the temperature dependence of $\Delta f$ at $17.1$\,nm$^{-2}$ either.}
   \label{fig-3He}
\end{figure}

Figure\,\ref{fig-Tvsdf}(f) depicts a density variation of the fitted $\zeta$ value.
It increases rather abruptly by 5\% at $\rho = \rho_{2}$, the beginning of the liquid+QLC coexisting region, where $\Delta f (30\,\rm{mK})$ decreases sharply. 
This change indicates an abrupt enhancement of the viscoelastic coupling between the He and graphite when a fraction of the QLC phase appears along the platelet edges in the Grafoil substrate attributed to stiffening driven by a structural change in the $^4$He film.
Then, the present results indicate that the $^4$He-QLC phase shows novel superfluidity with enhanced viscoelasticity comparable to that in the solid phase below 0.4--0.5\,K that is qualitatively different from ordinary frictionless superfluidity in the liquid phase.
If this interpretation is correct, a fermionic counterpart, the $^3$He-QLC phase, should also show a comparable increase in $\zeta$ but no superfluidity.
Indeed, as seen in Fig.\,\ref{fig-3He}, we observed a similar $\zeta$ increase in $^3$He to that in $^4$He at a density near the beginning of the liquid+QLC coexistence, while no superfluid response at any densities (inset).

Finally, let us consider why the standard BKT transition nature, i.e., a sharp $\Delta f$ jump at a well-defined $T_{\rm{BKT}}$, is not observed in this system.
This can be attributed to elementary~\cite{Nyeki2017a} or topological~\cite{Nakamura2016} excitations based on the proposed exotic ground states (intrinsic scenario).
However, it is well known that \textit{transport} measurements for low dimensional superconductors are sensitive to sample quality such as grain boundaries.
Therefore, similar precautions must be taken in TO measurements for 2D superfluids adsorbed on exfoliated graphite substrates, where microcrystallites (platelets; 10--20\,nm size~\cite{Birgeneau1982,Niimi2006}) are interconnected by grain boundaries (platelet edges).
Indeed, there are several indications that this extrinsic scenario must be seriously considered. 
First, the standard BKT transition is known to recover layer by layer from two to four layers of $^4$He films on Grafoil~\cite{Crowell1996,Usami2022}, suggesting diminishing substrate effects.
Most prominently, the density lag between $\rho_{\rm{1st}\rightarrow\rm{2nd}}$ and $\rho_{1}$ is 5.2\,nm$^{-2}$ in the second layer, while it is only 0--0.5\,nm$^{-2}$ in the third layer~\cite{Crowell1996}.
Second, the QLC phase shows a higher $T_{\rm{onset}}$ than the liquid phase, which contradicts the naive expectation from the intrinsic scenario.
It is also strange that both the liquid and QLC phases share the same $\log T$-dependence regardless of their different ground states.

\begin{figure}[b]
	\centering
	\includegraphics[width=1.00\linewidth]{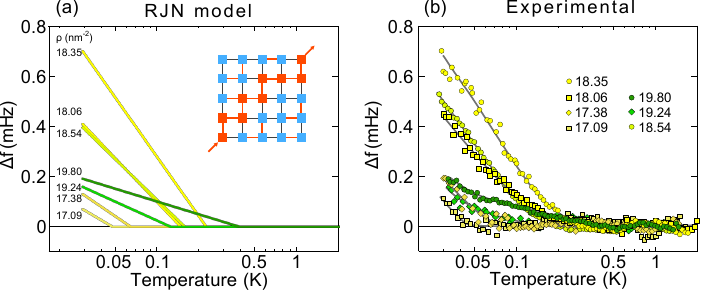} 
			\caption{(a) Temperature dependences of $\Delta f$ calculated from the RJN model. The magnitude of $\Delta f$ is normalized to the measured value for the 18.35\,nm$^{-2}$ sample at $T = 30$\,mK. (Inset) Schematic image of the RJN model. Superfluid small islands (square pads) are interconnected via active (red lines) or inactive (blue lines) Josephson junctions. The global coherence is supported by a percolated superfluid path shown in red. Note that, in the present model calculation, we have not assumed any particular lattice or coordination number such as the square lattice illustrated here. (b) Temperature dependences of experimental $\Delta f$ at selected densities (obtained from Figs.~\ref{fig-Tvsdf}(a)--(c)).}
	\label{fig-RJN}
\end{figure}

We thus developed the ``random Josephson network (RJN)'' model, extending the concept originally proposed in Ref.~\cite{Crowell1996}, to account for the behavior, particularly in the liquid phase.
We modeled our system as $N$ identical 2D superfluid nanoislands, each connected to others via a single most-conductive Josephson junction (see inset in Fig.~\ref{fig-RJN}(a)).
If the Josephson energy $e_{\rm{J}}$ of a junction with width $w$, length $l$, and barrier height $U$ exceeds the thermal energy, i.e., $e_{\rm{J}} \geq k_{\rm{B}}T$, the junction supports the superfluid phase coherence between the islands ($\textit{active}$); otherwise it is $\textit{inactive}$.
For simplicity, we assume a distribution only in $l$ with $dn/dl=$\,const. (uniform distribution), where $n$ is the areal number density of junctions, while fixing all other parameters ($w$, $U$, and $dn/dl$) at their mean values.
When the longest active-junction length $l_{\rm{max}} (\rho, T)$ exceeds a certain percolation threshold $l_{\rm{p}}$, a finite superfluid response ($\Delta f$) is expected as
\begin{equation}
\Delta f(T, \rho) \propto \lambda (\rho) \left[\ln{\frac{z \rho_{\rm{s}}}{\lambda (\rho)T}} - \frac{l_{\rm{p}}}{\lambda (\rho)}\right],
\label{eq-f(T)}
\end{equation}
where $z \equiv 2 \hbar^{2} w / (k_{\rm{B}} m)$ and $\rho_{\rm{s}}$ is the superfluid number density.
$\lambda$ is the penetration depth for a $^4$He atom of mass $m$ tunneling through the junction with the energy scale set by its chemical potential $\mu (\rho)$; thus $\lambda$ depends on $\rho$.
For further details of the RJN model, see Ref.~\cite{sm}.

Despite the bold simplification and assumptions, Eq.\,\eqref{eq-f(T)} captures the essential features of the present experimental results for the liquid phase remarkably well.
The model naturally reproduces the density lag before $\rho_{1}$, the logarithmic $T$-dependence of $\Delta f$ ($\propto a \ln{T/T_{\mathrm{onset}}}$), and the growth of $a$ with increasing density from $\rho_{1}$ through $\rho_{2}$.
In Figs.\,\ref{fig-RJN}(a) and (b), the fitting results of the RJN model are compared with our experimental results.
The fitting result is also shown by the red solid curve in Fig.\,\ref{fig-Tvsdf}(e).

Within the RJN model, the observed rapid suppression of $\Delta f$ above $\rho_{2}$ is driven by the appearance of the QLC phase with the enhanced viscoelastic response and limited superfluidity ($\rho_{\rm{s}} < \rho_{\mathrm{2nd}}$) near the platelet edges, which makes the junction network less conducting overall.
$\rho_{2}$ is, therefore, an extremely useful {\it in situ} density-calibration point for high-precision comparison among TO data by different groups~\cite{sm}.
After exploring the parameter space, we found that the RJN model can still be applicable to the higher-density regions semi-quantitatively, assuming $\theta \equiv \rho_{\rm{s}}/\rho_{\mathrm{2nd}} =$ 0.3--0.5 and a roughly factor-of-three reduction of $l_{\rm{p}}$ compared to that in the liquid phase, while keeping the density of states ($dn/dl$) unchanged.
The estimated upper limit of the superfluid fraction $\theta$ in the solid phase is 0.06, within which the supersolidity has not been observed in this system. 
As shown by the red dashed curve in Figs.\,\ref{fig-Tvsdf}(e) and \ref{fig-RJN}, the fitting quality is very good.
However, caution must be taken for the high-density fittings because we needed to introduce an additional assumption on the network configuration~\cite{sm}.

In summary, by performing simultaneous HC and TO measurements, we established a unified phase diagram that overlays superfluid response onto the well-established calorimetric phases of the second layer of $^4$He on graphite.
This enables an unambiguous phase-by-phase assignment of superfluidity while avoiding density ambiguities.
Owing to the moderate, well-characterized $T$- and $\rho$-dependences of the TO background frequency shift, we found that the proposed QLC phase intervening between the quantum liquid and solid exhibits reduced superfluidity below 0.5\,K together with enhanced viscoelasticity.
This strongly supports the superfluid liquid-crystal scenario~\cite{Nakamura2016}, an exotic superfluid coexisting with partial crystalline order (likely a hexagonal density modulation).
A phenomenological percolation model---a randomly distributed Josephson-junction network of 2D $^4$He nanoislands---accounts for the widely observed non-BKT onset of the superfluid response, including its $\log T$-dependence, suggesting a substrate origin.
These results motivate future TO, rheology or other transport measurements to resolve the intrinsic superfluidity of the QLC phase, using macroscopically uniform graphite substrates ~\cite{Nakamura2018}, mesoscopic detection techniques ~\cite{Todoshchenko2022}, and direct probes of its microscopic structure~\cite{Yamaguchi2022}.
The results will advance a unifying framework with direct implications for similar superfluids in more complex quantum systems.

{\it Acknowledgements---}
We thank Ryo Toda for his expert technical assistance, particularly in the TO measurements, and Sachiko Nakamura, Keiya Shirahama, and Yoshiyuki Shibayama for valuable discussions during the early stages of this work.
We are grateful to Department of Physics at the University of Tokyo for extended access to laboratory space, as well as Akira Yamaguchi and Akihiko Sumiyama for their kind hospitality at University of Hyogo, where part of this work was carried out.
This work was financially supported by Japan Society for the Promotion of Science (JSPS) KAKENHI Grant Number JP18H01170. 
J.U. was supported by JSPS through Program for Leading Graduate Schools (MERIT) and Grant-in-Aid for JSPS Fellows JP20J12304.
He is grateful to Satoshi Murakawa for insightful guidance throughout his graduate studies.


\bibliography{Ref}

\end{document}


\preprint{APS/123-QED}

\title{Supplemental Material for\\
On the Novel Superfluidity in the Second Layer of $^4$He on Graphite}

\author{Jun Usami}
\affiliation{Cryogenic Research Center, The University of Tokyo, 2-11-16 Yayoi, Bunkyo-ku, Tokyo 113-0032, Japan\looseness=-1}
\affiliation{Department of Physics, The University of Tokyo, 7-3-1 Hongo, Bunkyo-ku, Tokyo 113-0033, Japan\looseness=-1}
\affiliation{National Institute of Advanced Industrial Science and Technology (AIST), 1-1-1 Higashi, Tsukuba, Ibaraki 305-8565, Japan\looseness=-1}
\author{Hiroshi Fukuyama}%
\affiliation{Cryogenic Research Center, The University of Tokyo, 2-11-16 Yayoi, Bunkyo-ku, Tokyo 113-0032, Japan\looseness=-1}
\affiliation{Department of Physics, The University of Tokyo, 7-3-1 Hongo, Bunkyo-ku, Tokyo 113-0033, Japan\looseness=-1}

\date{\today}


\maketitle



%

\subsection{Phase assignment using heat-capacity data}
\label{Phase assignment}

When we quantitatively compare our HC data of 2D $^{4}$He samples with those by other workers, we have to normalize their data by multiplying by a constant factor $\sigma_{0}$ ($= A'/A''$), assuming negligibly small uncertainties in measuring He amounts and heat introduced into calorimeters.
Here $A'$ and $A''$ are the surface areas of our and their substrate, respectively.
The surface area of the graphite substrate in a calorimeter is determined by measuring the $^{4}$He amount ($n_{1/3}$) necessary for complete formation of the submonolayer $\sqrt{3}\times\sqrt{3}$ commensurate phase ($1/3$ phase), because its areal density is precisely known as 6.366\,nm$^{-2}$.
$n_{1/3}$ is measurable either through the ``sub-step'' in an N$_{2}$ adsorption isotherm at a fixed temperature around 74--77\,K~\cite{Nakamura2012a,Nakamura2016} or through the highest HC melting peak of $^{4}$He at 2.9\,K~\cite{Bretz1973,Greywall1993,Nakamura2013a}.
The signatures are sharp enough in both types of measurements, and the adsorption isotherm measurement can be carried out using the same calorimeter. 
Thus, in general, comparison among different HC measurements is much more reliable than comparisons among different physical quantities like $C$ vs. $\Delta f$.

Nonetheless, the determination of the surface area of exfoliated graphites is still subject to errors of the order of 2--8\% depending on the group, measurement temperatures, fraction of surface heterogeneities, accumulated small systematic errors, etc.~\cite{Lauter1990,Godfrin1995,Bauerle1996a,Leao2000}.
Thus, we fine-tuned the $\sigma$ parameter around $\sigma_{0}$ so as to obtain the best matching among HC data obtained in this work using the Grafoil substrate and the previous data reported by Greywall~\cite{Greywall1993} and Nakamura \textit{et~al}.~\cite{Nakamura2016}.
Note that the platelet size of ZYX is more than ten times larger than that of Grafoil. 
This enabled Nakamura \textit{et~al}.~\cite{Nakamura2016}, who used ZYX, to resolve the detailed and complicated phase diagram in this region (see Fig.\,\ref{fig-Recalibration}(e) or Figs.~\ref{fig-density}(a)(b)), which had not been fully understood in the data of Greywall~\cite{Greywall1993} who used Grafoil.

\renewcommand{\figurename}{Fig.}
\setcounter{figure}{0}
\renewcommand{\thefigure}{S\arabic{figure}}
\begin{figure}[h]
\centering
 \includegraphics[width=0.60\linewidth]{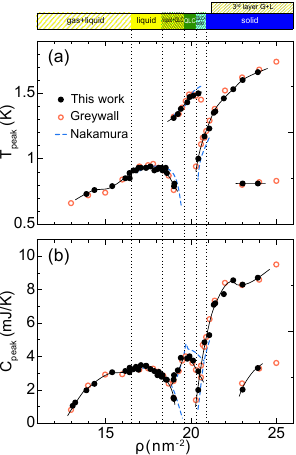}
    \caption{Density dependences of (a) $C_{\rm{peak}}$ and (b) $T_{\rm{peak}}$ obtained in this work (black dots), Ref.~\cite{Greywall1993} (open circles), and Ref.~\cite{Nakamura2016} (dashed curves) for the second layer of $^4$He on graphite. The substrate used in this work and Ref.~\cite{Greywall1993} is Grafoil, while that in Ref.~\cite{Nakamura2016} is ZYX. Here, the heat capacities measured in Refs.~\cite{Greywall1993} and \cite{Nakamura2016} are multiplied by $\sigma = 0.175$ and 1.50, respectively. The solid curves are guides to the eye. The thermodynamic phase diagram established in Ref.~\cite{Nakamura2016} is shown at the top. Note that the vertical dotted lines do not correspond to $\rho_{0}$--$\rho_{5}$ shown in Figs.\,3(e) and (f) but to the phase boundaries claimed in Ref.~\cite{Nakamura2016}.}
\label{fig-density}
\end{figure}

Figures~\ref{fig-density}(a) and (b) compare our HC data (black dots) with those from Ref.~\cite{Greywall1993} (open circles) and Ref.~\cite{Nakamura2016} (dashed curves), after the fine tuning of $\sigma$.
They depict density dependences of (a) the HC peak $C_{\rm{peak}}$ and (b) the peak temperature $T_{\rm{peak}}$. 
Excellent agreement with the data from Ref.~\cite{Greywall1993} within $\pm0.5$\% is achieved when $\sigma = 0.170(2)$.
This value is only slightly (by 3\%) smaller than $\sigma_{0} = 0.175(1)$ calculated from the stated surface area $A''=261$\,m$^{2}$ in Ref.~\cite{Greywall1993}, yet it provides clearly better matching.
The same is true for the comparison with the data from Ref.~\cite{Nakamura2016}, where the fine tuned $\sigma$ is 1.46(2).
This is again only slightly smaller (by 3\%) than $\sigma_{0} = 1.50(1)$ calculated from the stated $A''$ in Ref.~\cite{Nakamura2016}.
Based on these refinements, the density axis for our data in Fig.~\ref{fig-density} has also been corrected by $+3\%$, adopting the density scales in Refs.~\cite{Greywall1993,Nakamura2016}.
As is expected, this correction further improves the agreement among the three datasets.
This means that the HC data obtained by different groups can be scaled (or compared with each other) using only a single parameter $\sigma$ (ratio of the surface areas) with a $\pm0.5$\% accuracy.
Consequently, throughout the \textit{simultaneous} HC measurements, we are able to plot our TO data as a function of density, which is perfectly consistent (within $\pm0.5$\%) with the density scale of the currently most detailed and reliable thermodynamic phase diagram established in Ref.~\cite{Nakamura2016}.

The surface area of our Grafoil substrate was determined originally as $A' = 45.7(3)$\,m$^{2}$ within a nominal precision of $\pm0.7$\%, using the $1/3$-phase sub-step in N$_{2}$ adsorption isotherm as a calibration point~\cite{Usami2022}.
Nonetheless, we needed to multiply our original density scale by 1.03, that is, we rescale the surface area $A'$ to $A$ ($= 44.3$\,m$^{2}$), in order to be consistent with the density scale of Refs.~\cite{Greywall1993,Nakamura2016}.
The 3\% rescaling is within typical uncertainties for determination of the substrate surface area (2--8\%) as mentioned earlier, but, unfortunately, it is large enough to make it difficult to compare the $\Delta f$ results from non-simultaneous TO experiments with the thermodynamic phase diagram.

Next, based on our HC measurements, we discuss the reconfirmation of the absence of sharp HC anomalies that indicate first- or second-order melting transitions at densities near the QLC phase.
If commensurate solids such as the $4/7$ or $7/12$ phases do exist in the second layer, those sharp anomalies should be observed.
Instead, what is actually observed is quite broad melting HC anomalies around 1.4\,K, regardless of the substrate quality (no size effects).
Therefore, the existing HC data~\cite{Greywall1993,Nakamura2016}, including the present study, are in sharp contrast to most Monte Carlo predictions~\cite{Abraham1990,Pierce1998,Pierce1999a,Takagi2009,Gordillo2018,Ahn2016,Gordillo2020}, while some other studies suggest the absence of commensurate phases~\cite{Corboz2008,Moroni2019,Boninsegni2020}.
It is noted that the system size in these {\it ab initio} calculations is less than 400 atoms, on the other hand, it is of the order of 10$^{4}$ (Grafoil) and 10$^{6}$ (ZYX) atoms experimentally.

\subsection{Density rescaling in previous TO experiments}

It would be meaningful to compare the previous TO results after rescaling their density scales using $\rho_{2}$ ($=18.35$\,nm$^{-2}$) as a new calibration-point.
This is because the signature of $\Delta f$ at $\rho_{2}$ is very sharp and lies right within the density range of interest.

\begin{figure}[h]
\centering
 \includegraphics[width=0.80\linewidth]{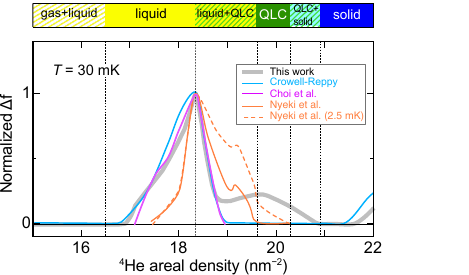}
\caption{Comparison of the superfluid responses $\Delta f$ in the second layer of $^4$He on Grafoil obtained by the previous TO experiments~\cite{Crowell1996,Nyeki2017a,Choi2021} at $T = 30$\,mK and ~\cite{Nyeki2017a} at 2.5\,mK. Each density scale was recalibrated by multiplying the original density axis by a constant factor so that the $\Delta f$ peak occurs at 18.35\,nm$^{-2}$ ($\rho_2$), the lowest density bound of the liquid+QLC coexistence region. The constant factors are 0.997, 1.014, and 1.007 for the data of Refs.~\cite{Crowell1996},~\cite{Nyeki2017a}, and~\cite{Choi2021}, respectively. The $\Delta f$ values of each dataset are normalized by its peak value. The result of the present work is also plotted as the thick gray curve. The reappearance of superfluidity above 21.5\,nm$^{-2}$ originates from the third layer. The top panel and the vertical dotted lines show the thermodynamic phase diagram determined in Ref.~\cite{Nakamura2016}.}
\label{fig-Recalibration}
\end{figure}

The results are shown in Fig.~\ref{fig-Recalibration}.
The highest density studied by Choi \textit{et al}.~\cite{Choi2021} using the rigid TO cell was likely around 18.9\,nm$^{-2}$ (on the density scale of Refs.~\cite{Greywall1993,Nakamura2016}) within the liquid+QLC coexistence region, not reaching the pure QLC phase.
Or, the subtle superfluid signal of the QLC phase with $T_{\rm{onset}}=$\,0.4--0.5\,K might have been neglected because of the too low background fitting-range ($T_{\rm{high}} = 0.5$\,K)~\cite{Choi2021}.
The TO experiment carried out by Ny{\'{e}}ki \textit{et al}.~\cite{Nyeki2017a} down to 2.5\,mK probably covered up to about 20.2\,nm$^{-2}$ near the high-density boundary of the pure QLC phase. 
Thus, the finite $\Delta f$ signals they observed well below 10\,mK for densities above 19.6\,nm$^{-2}$ likely correspond to the superfluid NCRI signals in the QLC phase identified in this work.
Another report by Ny{\'{e}}ki \textit{et al}.~\cite{Nyeki2017}, where the authors suggested the possible supersolidity in the pure solid phase, presumably covered only up to the QLC phase, not the solid phase.

The rescaled results from Refs.~\cite{Crowell1996,Choi2021} agree very well with our data at least below $\rho_{3}$ ($=18.8$\,nm$^{-2}$), as seen in Fig.~\ref{fig-Recalibration}.
The lack of superfluid detection between $\rho_{3}$ and $\rho_{4}$ ($=20.9$\,nm$^{-2}$) in Ref.~\cite{Crowell1996} is likely due to the relatively large uncertainties in their background subtraction above 300\,mK.

\subsection{Composite backgrounds in the previous TO experiments for various He samples}

The resonance frequency of TO is given by $2\pi f=\sqrt{\kappa/I}$, where $I$ and $\kappa$ are the moment of inertia mainly of the bob (sample chamber) on top of the torsion rod and the torsional modulus mainly of the torsion rod, respectively.
At low temperatures below 2\,K, where thermal contractions of materials are negligibly small, $I$ can vary with temperature either because of superfluid mass decoupling below 0.4--0.5\,K, which is of interest to us, or helium desorption from the substrate above 1.3--1.5\,K depending on the density.
The latter can be corrected reliably because it shows an exponentially rapid increase at higher temperatures.
On the other hand, $\kappa$ varies with temperature due to dislocation pinning/depinning in the torsion rod, hence its temperature dependence depends on the rod material~\cite{Nyeki2017a,Kleiman1985,Adams1991} and can be accurately determined by measuring the empty TO cell, $f_{\rm{emp}} (T)$.
However, it is known that $\kappa$ usually depends not only on the temperature but also on the sample density $\rho$, providing the substrate-adsorbate composite background, $f_{\rm{cBG}} (T, \rho)$~\cite{Maris2012,Makiuchi2018}.
When this composite effect is moderate, as in our TO, the $T$- and $\rho$-dependences are conveniently evaluated using Eq.\,(1) with $\zeta = 1.2$--$1.4$, as shown in the main text.

\begin{figure}[h]
\centering
 \includegraphics[width=0.90\linewidth]{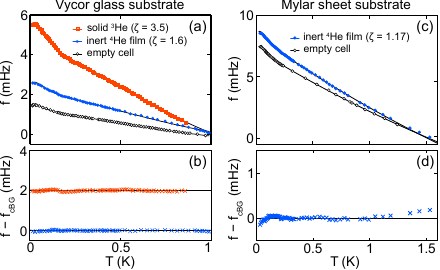}
       \caption{Temperature dependences of TO resonance frequencies $f(T)$ with and without He samples in the previous works, showing the substrate-sample composite effect. (a) The data for the TO cell used in Ref.~\cite{Kim2004} containing a Vycor substrate; empty cell (open circles), with an inert $^4$He film (blue closed circles), and with a bulk solid $^3$He at $P= 5.0$\,MPa (red open squares). The curves are fittings to Eq.\,(1) with $\zeta = 1$, 1.6, and 3.5, respectively. All data are plotted as shifts from the values at $T= 1$\,K. (b) Fitting residuals in (a). Each color corresponds to the color of the data in (a). (c)(d) Similar plots to (a)(b). Here the data obtained with the TO cell containing a Maylar substrate used in Ref.~\cite{Agnolet1989} are shown. The data with an inert $^4$He film (blue closed circles) are well fitted to Eq.\,(1) with $\zeta = 1.17$.}
\label{fig-comp_am}
\end{figure}

We point out that the simple empirical relation Eq.\,(1), which describes the composite (or viscoelastic) effects, is widely applicable to previous TO experiments regardless of the substrate material (either Vycor, Mylar or Grafoil) and He sample stiffness in a broad temperature range between 5\,mK and 1.6\,K.
The first example (Fig.~\ref{fig-comp_am}(a)) is the TO resonance frequency data from Ref.~\cite{Kim2004}, where the sample cell containing a Vycor glass substrate is loaded with a non-superfluid (or inert) $^4$He film (blue closed circles) and a bulk $^3$He solid at the pressure of 5.0\,MPa (red open squares).
The open symbols represent the empty-cell resonance frequency, $f_{\rm{emp}} (T)$, and all data are plotted as shifts from the value at $T= 1$\,K.
The basic variation with the temperature seen in these data is generally attributable to the stiffness variation of the BeCu torsion rod, while its magnitude is enhanced by the presence of the stiffer He samples.
The curves through the data points are composite backgrounds $f_{\rm{cBG}}$ obtained by least-squares fitting to Eq.\,(1) with $\zeta =1.6$ (inert He films) and $3.5$ (solid $^3$He) in the range of 20\,mK $\leq T \leq 1$\,K.
As has been already remarked by the author himself~\cite{Kim2004}, the fitting works very well (or surprisingly well even for such a large $\zeta$ value), as seen in Fig.~\ref{fig-comp_am}(b), where the deviations from the fits are plotted.

Another example is shown in Figs.~\ref{fig-comp_am}(c)(d).
Here, the cell contains a Mylar sheet substrate being loaded with an inert $^4$He film~\cite{Agnolet1989}.
Again, Eq.\,(1) represents the data for the $^4$He film with $\zeta= 1.17$ very well in the range of 5\,mK $\leq T \leq 1.6$\,K.
We note that similar applicability is found even for the more rigid TO cells used in Ref.~\cite{Crowell1996} ($\zeta = 1.06$) and Ref.~\cite{Choi2021} ($\zeta = 1.035$) (the data are not plotted here).

Therefore, the simple and linear composite effect represented by Eq.\,(1) appears to be widely applicable over a broad temperature range at least between 20\,mK and 1\,K, as long as the torsion rod is made of BeCu.
Note, however, that a more complicated and non-linear composite effect is reported for a TO cell with a torsion rod made of coin silver~\cite{Nyeki2017a}.

\subsection{Random Josephson network (RJN) model}

\begin{figure}[b]
\centering
 \includegraphics[width=0.85\linewidth]{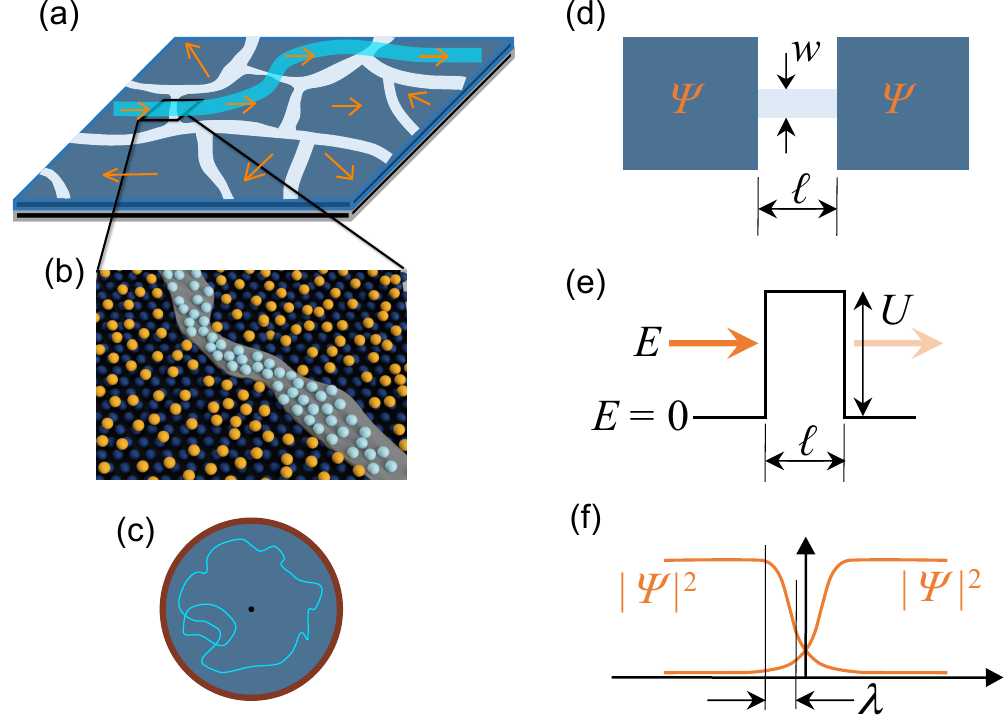}
\caption{(a) Schematic of atomically thin superfluid $^4$He films adsorbed on an exfoliated graphite substrate such as Grafoil. Many small 2D He islands are randomly interconnected with platelet edges (white curves). The arrows symbolize the order-parameter amplitudes and phases of the individual He islands. The blue curve indicates a superfluid path connecting some He islands through active Josephson junctions. (b) Schematic of the atomic arrangement near a platelet edge. The edge is preferentially covered by amorphous $^4$He with higher and lower densities than those in He islands. Among the many junctions existing along a single edge, we consider only the most strongly coupled one. (c) Schematic top view of the TO cell with a simply-connected macroscopic superfluid path (blue curve), which can be detected as an NCRI signal. (d) Two superfluid 2D He islands connected by an active junction. (e) Approximated potential barrier profile through which a $^4$He atom tunnels in (d). The critical mass current through the junction shown in (d) can be estimated considering this tunneling problem. (f) Overlap of the wave function amplitudes of the islands through the junction with a penetration length of $\lambda$.}
\label{fig-RJN1}
\end{figure}
\begin{figure*}[t]
\centering
 \includegraphics[width=0.85\linewidth]{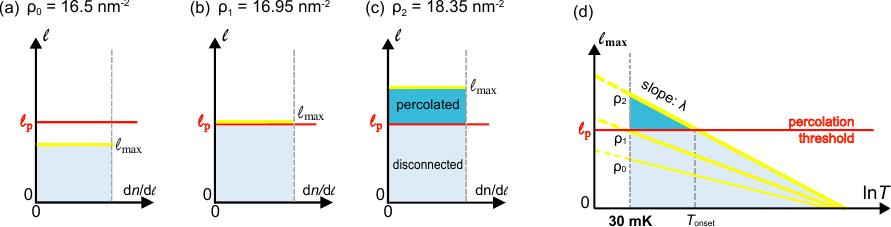}
\caption{Schematic diagrams explaining the RJN model for the liquid phase in the second layer of $^4$He on graphite. (a)--(c) Distributions of the junction length $l$ at $\rho =$ (a)$\rho_0$, (b)$\rho_1$, and (c)$\rho_2$ at $T=30$\,mK. The constant density of states ($dn/dl$) for $l$ is assumed here. The colored areas below $l_{\rm{max}}$, the maximum active junction length (yellow lines), correspond to active junctions. Only active junctions with $l$ longer than the percolation threshold length $l_{\rm{p}}$ (red lines) contribute to the superfluid responses in TO experiment (dark blue area denoted by ``percolated''). Active junctions with $l < l_{\rm{P}}$ are not involved in the simply connected superfluid paths (light blue area denoted by ``disconnected'') in (c). (d) Temperature dependences of $l_{\rm{max}}$ calculated from Eq.\,\eqref{eq-lmax} at $\rho = \rho_0$, $\rho_1$, and $\rho_2$ (the yellow lines). The superfluid response $\Delta f$ observed in the TO experiment above 30\,mK is proportional to the dark blue area, cf. Eq.\,\eqref{eq-fitting}.
}
\label{fig-RJN2}
\end{figure*}

Regarding the well-known patchwork morphology of the Grafoil substrate consisting of microscopic graphite platelets, typically 10--20\,nm in size~\cite{Birgeneau1982,Niimi2006} (see Fig.~\ref{fig-RJN1}(a)), care must be taken when measuring transport properties of its adlayers.
Each platelet has an atomically flat surface, hence each 2D He adlayer is a clean 2D system of finite size.
However, the platelet edges are likely covered by non-superfluid He atoms of high- and low-density amorphous forms as illustrated in Fig.~\ref{fig-RJN1}(b).
In order for the NCRI signal to be detected in a TO measurement, global superfluid coherence must be established across the 2D He islands via many weak links at the edges (blue paths in Figs.~\ref{fig-RJN1}(a)(c)). 
This is analogous to electrical-resistance measurements on superconducting granular films, for which the model considering large arrays of superconducting weak links (Josephson junctions) was developed by Abraham~\textit{et~al}~\cite{Abraham1982}.
Inspired by their model, we have developed the RJN model specific to our problem.

Fig.~\ref{fig-RJN1}(d) illustrates two superfluid islands with an equal order parameter (macroscopic wave function) amplitude $|\Psi|$ each connected by a narrow channel (weak link or Josephson junction) of width $w$ and length $l$.
The critical superfluid mass-current $j_{\rm{c}}$ through this junction can be modeled as a tunneling problem for a free particle of mass $m$ and energy $E$ through a potential barrier of height $U$ (Fig.~\ref{fig-RJN1}(e)) (see also Eq.\,(64) of Ref.~\cite{Varoquaux2015}).
Converting the 3D problem discussed in Ref.\,\cite{Varoquaux2015} to the 2D case, replacing $E$ by the chemical potential $\mu$ of 2D $^4$He, and using $|\Psi|^{2} = \rho_{\rm{s}}$, we obtain 
\renewcommand{\theequation}{S\arabic{equation}}
\begin{eqnarray}
j_{\rm{c}} = (2 \hbar \kappa_{\rm{s}} w / m \lambda) e^{-l / \lambda},
\label{eq-j_s}
\end{eqnarray}
and
\begin{eqnarray}
\lambda = \hbar / \sqrt{2 m(U - \mu)}
\label{eq-lambda}
\end{eqnarray}
in the weak coupling limit ($l/\lambda > 1$).
$\kappa_{\rm{s}}$ ($= m\rho_{\rm{s}}$) is the superfluid mass density.
Note that, at temperatures of interest ($30 \leq T \leq 500$\,mK), $\rho_{\rm{s}}$ and $\lambda$ are expected to be $T$-independent and close to their corresponding values at $T=0$, because $T_{\rm{onset}} \ll T_{\rm{BKT}}$\,($> 800$\,mK).

In reality, the junction parameters $l$, $w$, and $U$ should be distributed over wide ranges in a complicated manner.
In our RJN model, however, we simplify the real system by considering only the uniform distribution of $l$ ($dn/dl =$\,const.), and assume that such a distribution holds at least within the temperature range we are dealing with (only over a factor of three variation in $l$).
We also assume that the superfluid response $\Delta f$ is proportional to the number of He islands $\it{simply}$ percolated by active junctions ($l \leq l_{\rm{p}}$) like Fig.\,\ref{fig-RJN1}(c).
Here, an active junction means $e_{\rm{J}} = \hbar j_{\rm{c}} / m \geq k_{\rm{B}}T$.
Then, $\Delta f$ is in proportion to the number of active junctions.
Under these simplifications and assumptions, the model predicts the following relations: 
\renewcommand{\theequation}{S\arabic{equation}}
\begin{eqnarray}
\Delta f(\rho, T) =  \frac{\eta \rho_{\rm{s}} g} {N} \frac{dn}{dl} \left[ l_{\rm{max}}(\rho,T) - l_{\rm{p}} \right],
\label{eq-fitting}
\end{eqnarray}
\begin{eqnarray}
l_{\rm{max}}(\rho,T) = \lambda (\rho) \left[ \ln{\frac{z \rho_{\rm{s}}}{\lambda (\rho)}} - \ln{T} \right],
\label{eq-lmax}
\end{eqnarray}
where $z = 2 \hbar^{2} w / (k_{\rm{B}} m)$ and $g = - (df/d\rho)_{{T = 1\,\rm{K}}}$.
The $g$ value is determined as $10.77(5)$\,Hz\,nm$^{2}$ from the mass loading data at $T = 1$\,K.
Eq.\,(2) is a simplified expression for Eqs.\,\eqref{eq-fitting}\eqref{eq-lmax}.
$\eta$ in Eq.\,\eqref{eq-fitting} is the connectivity of platelets specific to each graphite substrate, denoting the areal fraction of platelets participating in the NCRI response.
$\eta$ is assumed to be independent of $\rho$.
The tortuosity $\chi$ ($= 1 - \eta$) is a similar quantity more frequently used in previous TO studies.
The $\chi$ value of our Grafoil substrate is $0.94$, which is typical among $0.942$~\cite{Nyeki2017a} and $0.98$~\cite{Crowell1996,Choi2021} in the previous experiments.

From Eqs.\,\eqref{eq-fitting}\eqref{eq-lmax}, the following expressions for $T_{\rm{onset}}$ and $a$ are obtained:
\begin{eqnarray}
T_{\rm{onset}}(\rho) = \frac{z \rho_{\rm{s}}}{\lambda (\rho)} \exp{ \left[- \frac{l_{\rm{p}}}{\lambda (\rho)} \right]},
\label{eq-Tonset}
\end{eqnarray}
\begin{eqnarray}
a(\rho) =  \frac{\eta \rho_{\rm{s}} g} {N} \frac{dn}{dl} \lambda (\rho).
\label{eq-a}
\end{eqnarray} \\

\noindent{\bf {RJN model for the liquid phase}}\\
Figures~\ref{fig-RJN2}(a)--(d) schematically show how the RJN model works in the uniform liquid phase ($\rho_{\rm{0}} < \rho < \rho_{\rm{2}}$).
(a)--(c) illustrate the occupation of the density of states $dn/dl$ by active junctions, while (d) shows temperature variations of $l_{\rm{max}}$.
At $\rho < \rho_{\rm{1}}$, where $l_{\rm{max}}$ \,($\rho$,\,30\,mK)\,$< l_{\rm{p}}$, no simply connected superfluid path exists, and $\Delta f$ does not appear at least above $30$\,mK (see Fig.~\ref{fig-RJN2}(a)).
Only at $\rho = \rho_{\rm{1}}$, where $l_{\rm{max}}$\,($\rho$,\,30\,mK)\,$ = l_{\rm{p}}$, simply connected superfluid paths just appear (Fig.~\ref{fig-RJN2}(b)).
This explains the observed density lag of $\Delta f$ between $\rho_{\rm{0}}$ (or $\rho_{\rm{1st}\rightarrow\rm{2nd}}$) and $\rho_{\rm{1}}$.
At $\rho > \rho_{\rm{1}}$, where $l_{\rm{max}}$\,($\rho$,\,30\,mK)\,$ > l_{\rm{p}}$, a finite $\Delta f$ ($= -a\ln{T/T_{\mathrm{onset}}}$) corresponding to the blue regions in Figs.~\ref{fig-RJN2}(c) and (d) is observed below $T_{\rm{onset}}$.

\begin{figure}[t]
\centering
\includegraphics[width=0.75 \linewidth]{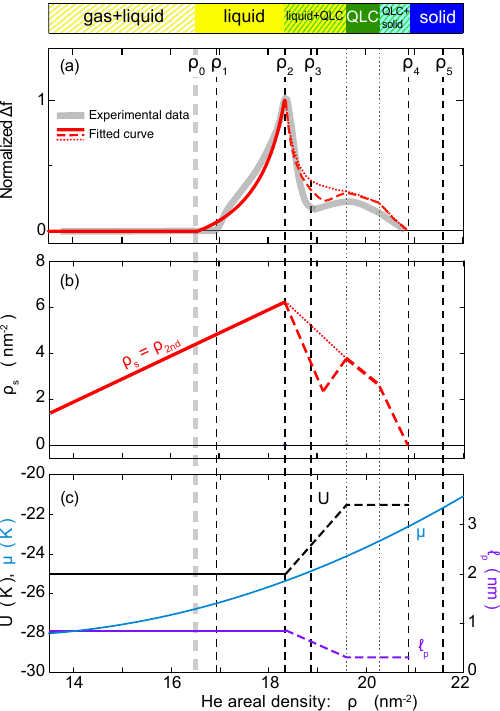}
\caption{Fitting results based on the RJN model and the corresponding fitting parameters for the superfluid response $\Delta f$ data at $T = 30$\,mK obtained in the present TO measurement on the second layer $^{4}$He on graphite. (a) Fitting results (red curves) and experimental data for $\Delta f$ (gray curve). Both are normalized to the maximum values at $\rho_2$. Eq.\,\eqref{eq-coex1} is used for the dotted curve, while Eqs.\,\eqref{eq-lmax2} and \eqref{eq-coex2} are used for the dashed curve, which provides a better fit in the liquid+QLC coexistence region. (b) Fitted superfluid number densities $\rho_{\rm{s}}$. The meanings of the dotted and dashed lines are the same as those in (a). The dashed lines in the liquid+QLC coexistence region represents $\rho_{\rm{s}}^{\rm{junction}}$ (see text). (c) Other fitting parameters: the potential barrier height $U$ (black line) and the percolation length limit $l_{\rm{p}}$ of the Josephson junctions (purple line). Also plotted is the theoretical chemical potential $\mu$ (blue line)~\cite{Gordillo2020}. The top panel shows the thermodynamic phase diagram determined in Ref.~\cite{Nakamura2016}.}
\label{fig-RJN3}
\end{figure}

Based on Eqs.\,\eqref{eq-lambda}, \eqref{eq-Tonset}, and \eqref{eq-a}, the model predicts that $a$ and $T_{\rm{onset}}$ increase with increasing $\rho$, because the chemical potential $\mu$ (and hence $\lambda$) should increase monotonically in the liquid phase.
These predictions are qualitatively consistent with the experimental observations in the present and previous works~\cite{Crowell1996,Nyeki2017a,Choi2021}.
Moreover, by using the chemical potential $\mu$ of second-layer liquid $^4$He on graphite calculated by Gordillo and Boronat based on the $\it{ab~initio}$ method~\cite{Gordillo2020} (blue solid curve in Fig.~\ref{fig-RJN3}(c)), it is possible to fit the entire density and temperature dependences of $\Delta f$ measured in the liquid phase even quantitatively (see Fig.\,6 and the red solid curves in Figs.\,3(e) and \ref{fig-RJN3}(a)).
This fitting was obtained with $U = -25$\,K and $l_{\rm{p}} = 0.85$\,nm, as plotted in Fig~\ref{fig-RJN3}(c).
The parameter $(dn/dl)/N$ was normalized based on the experimental $\Delta f (\rho_{2}, 30\,\rm{mK})$ value, assuming $\rho_{\rm{s}} = \rho_{\mathrm{2nd}} (\theta = 1)$.
Therefore, the RJN model captures the essential features of the superfluid response in the liquid phase below $\rho_{2}$. \\

\noindent{\bf RJN model for the QLC phase}\\
Above $\rho_{2}$, once the system enters the coexistence regime with the QLC phase, the nature of the Josephson network is expected to be modified in various ways.
Indeed, the experimental observations in the pure QLC phase, i.e., the increased $T_{\rm{onset}}$ and the decreased $a$ compared with those in the liquid phase, conflict with each other as intrinsic properties, and cannot be explained within the RJN model without increasing $U$ and reducing $l_{\rm{p}}$.
Then, we have tried to fit the data of the pure QLC phase by setting $U$ and $l_{\rm{p}}$ as free parameters, and obtained the reasonable result with $U = -20.4$\,K and $l_{\rm{p}} = 0.4$\,nm consistent with most of the experimental results, except that the calculated $\Delta f$ was larger than the measured value by about a factor of two.
Subsequently, we performed fitting by treating $\rho_{\rm{s}}$ as a free parameter as well, and obtained full agreement with the experimental data in the pure QLC phase with $U = -21.5$\,K, $l_{\rm{p}} = 0.3$\,nm, and $\theta =$ 0.3--0.5.

These modifications of the network parameters can be understood within the RJN model by the following two competing effects.
The appearance of the QLC phase, which has enhanced viscoelasticity and reduced superfluidity, in the close vicinity of the junctions increases the tunnel barrier $U$, thus reducing the conductivity of the Josephson network.
The concurrent compression of the first layer $^{4}$He~\cite{Lauter1991}, most likely during the liquid+QLC coexistence, reduces the percolation threshold $l_{\rm{p}}$, which makes the Josephson network more robust against thermal fluctuations.

The large reduction in $\theta$ is exotic but would be characteristic of superfluids with some degrees of crystalline order.
Similar reductions have been reported in possible supersolid systems, such as the hypothetical $7/12$ commensurate solid in the second layer of $^{4}$He on graphite ($\theta \approx 0.3$)~\cite{Gordillo2020} and 1D dipolar supersolids ($0.1 \leq \theta \leq 0.8$)~\cite{Biagioni2024}.

Here, $\mu (\rho)$ above $\rho_{2}$ was approximated as the blue curve in Fig.\,\ref{fig-RJN3}(c), indicating a smoothed interpolation between the calculated $\mu (\rho)$ for the liquid phase and that for the solid phase~\cite{Gordillo2020}.
This approximation would not significantly change the fitting results, because in the RJN model the important quantity is the difference between $U$ and $\mu$ rather than their absolute values (see Eq.\,\eqref{eq-lambda}. \\

\noindent{\bf RJN model for the coexistence density regions}\\
To fit the data in the liquid+QLC ($\rho_2 \leq \rho \leq$ 19.6\,nm$^{-2}$) and QLC+solid ($20.3 \leq \rho \leq \rho_{4}$) coexistence regions, we assumed:
\begin{eqnarray}
\rho_{\rm{s}} &= \rho_{\rm{s,1}}\sigma_1 + \rho_{\rm{s,2}}\sigma_2,
\label{eq-coex1}
\end{eqnarray}
where $\rho_{\rm{s,1}}$ ($\rho_{\rm{s,2}}$) is the superfluid number density in the coexisting phase-1(-2), and $\sigma_1$ ($\sigma_2$) is the areal fraction of the phase-1(-2) ($\sigma_1 + \sigma_2 = 1$).
Also, for simplicity, $U$ and $l_{\rm{p}}$ are assumed to vary linearly with $\rho$ connecting the values at the phase-1 and -2.
Note that $\rho_{\rm{s}} = 0$ in the solid phase from our TO measurements.
Within the QLC+solid coexistence region, we did not change the network parameters ($U$ and $l_{\rm{p}}$) because no significant change in $\zeta$ is detected there.
The fitted result for the whole density range above $\rho_{2}$ is shown as the red dotted curve in Fig.\,\ref{fig-RJN3}(a).
The result represents the overall feature of the experimental data (gray curve) fairly well, but does not reproduce the liquid+QLC coexistence region well.

The fitting quality in this region can be improved by treating the two $\rho_{\rm{s}}$ parameters involved in Eqs.\,\eqref{eq-fitting} and \eqref{eq-lmax} independently.
That is, we use the following expressions instead of Eqs.\,\eqref{eq-lmax} and \eqref{eq-coex1}:
\begin{eqnarray}
l_{\rm{max}}(\rho,T) = \lambda (\rho) \left[ \ln{\frac{z \rho_{\rm{s}}^{\rm{junction}}}{\lambda (\rho)}} - \ln{T} \right],
\label{eq-lmax2}
\end{eqnarray}
\begin{eqnarray}
\rho_{\rm{s}}^{\rm{junction}} &= \max \{ \rho_{\rm{s,1}} \sigma_{1} , \rho_{\rm{s,2}}\sigma_{2} \},
\label{eq-coex2}
\end{eqnarray}
while Eq.\,\eqref{eq-coex1} is still used in the prefactor of Eq.\,\eqref{eq-fitting}.
The physical meaning of this treatment is to select the most conductive junction for each He island stochastically, based on the principles of the RJN model.
The final fitting results obtained by this treatment are shown as the red dashed curves in Figs.\,3(e) and \ref{fig-RJN3}(a).
The resultant fitting parameters are shown in Figs.~\ref{fig-RJN3}(b)(c).
They reproduce our experimental TO results surprisingly well over the entire density range.
Nonetheless, we again emphasize that the present model may oversimplify the real complicated system, and that the fits at $\rho>\rho_{2}$ should be regarded as semi-quantitative.

\bibliography{Ref}